\documentclass{elsart}
\usepackage{epsf}
\begin{document}

\begin{frontmatter}
\title{Bohmian description of a decaying quantum system }
\author[McMaster]{Y. Nogami},
\author[Kyoto]{F.M. Toyama}, 
\author[McMaster,Redeemer]{W. van Dijk\thanksref{wvd}}
\address[McMaster]{Department of Physics and Astronomy, McMaster
University, Hamilton, Ontario, Canada L8S 4M1}
\address[Kyoto]{Department of Communication and Information 
Sciences, Kyoto Sangyo University, Kyoto 603-8555, Japan }
\address[Redeemer]{ Redeemer College, Ancaster, Ontario,
Canada L9K 1J4} 
\thanks[wvd]{E-mail: vandijk@physics.mcmaster.ca}

\begin{abstract}
We present a Bohmian description of a decaying quantum system. 
A particle is initially confined in a region around the origin 
which is surrounded by a repulsive potential barrier. The 
particle leaks out in time tunneling through the barrier. 
We determine Bohm trajectories with which we can visualize 
various features of the decaying system.
\end{abstract}
\begin{keyword}
 Bohm trajectories, $\alpha$-decay 
\end{keyword}
\end{frontmatter}

\section{Introduction}
Consider a nonrelativistic particle in a given potential field. In
Bohm's ontological interpretation of quantum mechanics, the position
and velocity of the particle are both well-defined at any instant of
time. The particle moves along a Bohm trajectory, irrespective of 
whether or not the particle is observed \cite{1,2,3,4}. The 
Schr\"{o}dinger wave function accompanies and guides the particle
\cite{5}. We do not talk about the collapse of the wave function 
caused by observation. A certain ``surrealistic" aspect of 
the Bohm trajectory in the presence of a Welcher Weg (which way) 
detector has been discussed recently but we do not consider such 
a situation in this Letter \cite{6}. 

The particle obeys Newton's equation of motion with a potential 
which consists of the usual given potential and the
``quantum potential". The latter is related to the amplitude of the
wave function. The initial condition of the motion, however, can be
determined only statistically. With this statistical uncertainty
implemented, Bohm's theory agrees with the traditional quantum
mechanics for all observable quantities. But the picture that
Bohm's theory offers is strikingly different from the traditional
one. A variety of problems have been examined on the basis of Bohm's
theory. Among them let us mention the recent work by Leavens et
al. and Oriols et al.~\cite{7,8} on the one-dimensional 
tunneling problem, which has direct relevance to what we propose to 
examine. The purpose of this Letter is to extend the application of 
Bohm's theory to the decay problem. 

We consider a model with a particle in a central potential $V(r)$. The
potential has a repulsive barrier that supports one or more unstable
bound states or resonances. The particle is initially confined inside
the potential barrier and at a certain time, $t=0$, it begins to leak
out. The model and its variants can be used to simulate a decay
process through tunneling such as the nuclear $\alpha$ decay~\cite{9}
or the emission of an electron from an artificial atom (quantum
dot)~\cite{9a}.  In order to apply Bohm's theory to such a system, one
has to know the wave function of the system explicitly. Bohm's
approach has not been attempted for the decay problem so far.  This is
because very little has been known about the wave function of such a
decaying system, particularly outside the potential barrier.  A
recently developed technique, however, has made it possible to solve
the time-dependent Schr\"{o}dinger equation accurately, at least for
the type of model that we consider below, no matter how large $r$ and
$t$ are \cite{10}.

In Section 2 we set up the model and we present a solution of 
the time-dependent Schr\"{o}dinger equation of the model with 
an appropriate initial condition. In Section 3 we work out the 
Bohmian description of the decay process and discuss 
various features such as the exponential decay law and deviation 
from it at very small time. A summary is given in Section 4.

\section{Model}
We assume a simple model with the repulsive potential barrier,
\begin{equation}
V(r) = (\lambda/a)\delta(r-a),
\label{1}
\end{equation}
where $\lambda >0$ and $a>0$ are constants. A particle is 
initially confined inside the potential barrier and it begins 
to leak out at $t=0$. This model has been used by a number of 
authors to examine various features of the decaying quantum 
system such as the exponential decay law and deviations from it 
at very small time as well as at very large time \cite{11,12,13}.
We use units such that $\hbar =1$ and $2m=1$ where $m$ is the 
mass of the particle of the model. In numerical illustrations 
we set $a=1$. For the strength of the potential we take 
$\lambda=6$, which is one of the cases assumed in earlier 
work~\cite{11,13}. The $\lambda$ corresponds to the $G$ 
of \cite{13}.

We consider only the $S$ state. The wave function $\psi (r,t)$
(actually the wave function times $r$) is determined by the
time-dependent Schr\"{o}dinger equation for $t>0$,
\begin{equation}
\mathrm{i}\frac{\partial \psi (r,t)}{\partial t}
=\left[ -\frac{\partial ^2}{\partial r^2} + V(r)\right]
\psi(r,t), \hspace{0.1in} \psi(0,t)=0.
\label{2}
\end{equation}
For the initial condition for $\psi (r,t)$, let us assume the 
normalized function
\begin{equation}
\psi(r,0) = \sqrt{\frac{2}{a}}
\sin \left(\frac{\pi r}{a}\right) \theta (a-r),
\label{3}
\end{equation}
where $\theta(x) = 1$ (0) if $x>0$ ($x<0)$. 

Although the model is very simple, its time-dependent Schr\"{o}dinger
equation is nontrivial to solve~\cite{11}. The wave function of the
model in the entire space was found in an analytical form only
recently~\cite{10}.  It reads as
\begin{equation}
\psi (r,t) = \sum _\nu c_\nu  
[M(k_\nu,r-a,t) + N_{-}(k_\nu,r-a,t)],
\label{4}
\end{equation}
\begin{equation}
N_{\pm}(k,x,t) = \frac{\mathrm{i}\lambda}{2ka}
[M(k,x,t)\pm M(k,-x,t)]\theta (-x), 
\label{5}
\end{equation}
where the summation is over $\nu=\pm 1, \pm 2, \ldots \,$.  The
$k_\nu$'s, which are determined by solving the equation 
$ka \cot ka + \lambda-\mathrm{i}ka = 0$ for $k$, 
are the positions of the poles of the {\bf S} matrix. They are 
all in the lower half of the complex $k$ plane. We designate the 
poles in the fourth (third) quadrant with $\nu = 1,2, \ldots \,$ 
($\nu = -1,-2,\ldots$). For $r>a$, $c_\nu = -2\pi \mathrm{i} a_\nu (r)
\mathrm{e}^{\mathrm{i}k_\nu a}$ for the $a_\nu(r)$ defined
in~\cite{10}. It is given by
\begin{equation}
c_\nu = \frac{2\pi \sqrt{2a}k_\nu}{(k_\nu^2 a^2-\pi^2)
[(1+\lambda-\mathrm{i}k_\nu a)\cot k_\nu a - \mathrm{i}-k_\nu a]}.
\label{6}
\end{equation}	
The $c_\nu$'s satisfy $\sum _\nu c_\nu /k_\nu=0$, but
$\sum _\nu c_\nu \neq 0$.
The $M(k,x,t)$ is the Moshinsky function \cite{14}
\begin{equation}
M(k,x,t) =\frac{1}{2} \mathrm{e}^{-\mathrm{i}k^2 t}
\mathrm{e}^{\mathrm{i}kx}
{\rm erfc}(y),
\hspace{0.05in}
y = \mathrm{e}^{-\mathrm{i}\pi/4}\left(\frac{x-2kt}{2\sqrt{t}}\right),
\label{7}
\end{equation}
where ${\rm erfc}(y) = (2/\sqrt{\pi})\int ^\infty _y \mathrm{e}^{-u^2} 
\mathrm{d}u$.  
In the limit of $t \rightarrow 0$, $M(k,x,t)$ becomes discontinuous 
at $x=0$. Otherwise $M(k,x,t)$ is a smooth function of $x$.  The
$\psi(r,t)$ for $t>0$ is continuous at $r=a$ as can be seen from
$N_{-}(k,0,t)=0$.

The $r$-derivatives of $\psi (r,t)$ are given by
\begin{eqnarray}
\psi '(r,t)&=&\mathrm{i}\sum _\nu c_\nu \{k_\nu
[M(k_\nu,r-a,t) \nonumber \\
&+& N_{+}(k_\nu,r-a,t)] + \chi (r-a,t) \},
\label{8}
\end{eqnarray}
\begin{eqnarray}
&& \psi ''(r,t) = \sum _\nu c_\nu \{-k_\nu ^2 [M(k_\nu,r-a,t)+
N_{-}(k_\nu,r-a,t)] \nonumber \\ 
&& \hspace{.3in} +\frac{r-a+2k_\nu
t}{2t}\chi(r-a,t)\} + \frac{\lambda}{a}\delta (r-a)\psi (a,t),
\label{9}
\end{eqnarray}
\begin{equation}
\chi(x,t)= \frac{\mathrm{e}^{\textstyle \mathrm{i}\pi/4}}{2\sqrt{\pi t}}
\exp {\left(\frac{\mathrm{i}x^2}{4t}\right)}.
\label{10}
\end{equation}
It is not difficult to confirm that $\psi$ of Eq.~(\ref{4}) does 
satisfy Eq.~(\ref{2}). The $\delta$-function part of $\psi''$
exactly cancels $V\psi$ in the Schr\"{o}dinger equation.
In deriving $\psi$, $\psi'$ and $\psi''$ we used 
$\sum _\nu c_\nu /k_\nu =0$. The convergence of the $\nu$ 
summation for $\psi$ can be dramatically improved by adding
$\chi(r-a,t)\sum_\nu c_\nu/k_\nu$, which is formally zero, to
the right hand side of Eq.~(\ref{4}). 

\section{Bohmian description}
We write the wave function as
\begin{equation}
\psi (r,t) = R(r,t) \exp{[\mathrm{i}S(r,t)]},
\label{11}
\end{equation}
where $R(r,t)$ and $S(r,t)$ are both real. Then Eq.~(\ref{2})
leads to
\begin{equation}
\frac{\partial S}{\partial t} + (S')^2 + U(r,t) = 0,
\label{12}
\end{equation}
\begin{equation}
U(r,t) = V(r) + Q(r,t), \hspace{0.1in} Q(r,t) = - R''/R, 
\label{13}
\end{equation}
\begin{equation}
\frac{\partial R^2}{\partial t} + 2(R^2 S')' = 0,
\label{14}
\end{equation}
where $S'=\partial S/\partial r$ and 
$R'' = \partial ^2 R /\partial r^2$. The $Q(r,t)$ is the 
quantum potential. 

In Bohm's interpretation, the particle obeys Newton's equation of
motion with the potential $U(r,t)=V(r)+Q(r,t)$. Equation~(\ref{12}) 
is the classical Hamilton-Jacobi equation with potential $U(r,t)$. 
The momentum $p=mv$ ($m=1/2$) of the particle at $(r,t)$ is given 
by
\begin{equation}
p = mv = S'(r,t).
\label{15}
\end{equation}
This $p$ should not be confused with the usual quantum momentum 
operator.  Equation (\ref{15}) together with an appropriate 
initial condition determines the particle trajectory. The motion 
is causal and deterministic. The underlying potential $U(r,t)$ 
is of a nonlocal and holistic nature.  It depends on the wave 
function which in turn is related to aspects of the system at 
points different from $r$.  

The motion is subject to uncertainty in the sense that the 
initial condition is known only statistically. The particle may 
start at any point $r=r_0$ within the potential barrier; 
$0<r_0<a$. Since $S(r,0)=0$ for the $\psi(r,0)$ of Eq.~(\ref{3}), 
the initial velocity is zero, i.e., $v(r_0,0)=0$. Two trajectories 
starting at different points at $t=0$ do not cross each other 
\cite{2,7,8}\footnote{ Since we are considering only the $S$ 
state the wave function is independent of angles. The velocity
associated with a Bohm trajectory has only a radial component.  The
particle that starts at $r_0$ moves always in the radial direction. In
this Letter by a trajectory we mean a plot of $r$ versus $t$, not a
path in the $xyz$ configuration space.}.  It is understood that the
probability with which the particle starts at $r_0$ is proportional to
$|\psi(r_0,0)|^2$. An observable quantity of the system is calculated
by taking an average over all possible trajectories weighted according
to $|\psi (r_0,0)|^2$. Equation (\ref{14}) guarantees the conservation
of the probability of the particle in the statistical ensemble of the
trajectories.

With the wave function given in Section 2 we can determine $R$, 
$S$ and their derivatives by using ${\rm Im}(\psi ' /\psi)=S'$, 
${\rm Re}(\psi '' /\psi)=(R''/R)-{S'}^2$, etc. At $r=a$, it 
can be shown that $R$, $S$ and $S'$ are all continuous, while 
$ R'(a+0,t) - R'(a-0,t)= (\lambda/a)R(a,t)$. The quantum 
potential $Q$ contains a $\delta$-function term, which cancels 
the $\delta$-function of $V$ in potential $U$.  When $t>0$, 
$U$ is continuous at $r=a$. At $t=0$, we have
\begin{equation}
Q(r,0) = \left(\frac {\pi}{a}\right)^2, \hspace{0.1in} r<a.
\label{16}
\end{equation}
Unlike for $t>0$, there is no exact cancellation of the 
$\delta$-function terms of $Q$ and $V$ at $t=0$ and hence $U(r,0)$ 
is singular at $r=a$. As soon as $t$ becomes positive, this 
singularity disappears. Let us define the energy $E(r,t)$ of the 
particle by
\begin{equation}
E(r,t) = p^2 + U(r,t).
\label{17}
\end{equation}
The particle can start at any point $0<r_0<a$. Since $v(r_0,0)=0$, 
the initial value of the energy is $E(r_0,0)=(\pi/a)^2$ which is
independent of $r_0$.  Note, however, $E(r,t)$ is not conserved 
during the motion because $U(r,t)$ depends on $t$ explicitly.

The trajectories can be labeled in terms of the starting point
$r_0$. Instead of $r_0$, it is often more convenient to use
\begin{equation}
s(r_0) = \int_0 ^{r_0}\psi ^2 (r,0) \mathrm{d}r
=\frac{r_0}{a} - \frac{1}{2\pi}
\sin \left(\frac{2\pi r_0}{a}\right),
\label{18}
\end{equation}
which is the probability for the particle being in the region
$(0,r_0)$ at $t=0$. Note that $s(a)=1$, $s(a/2)=1/2$ and $s(0)=0$.
For the trajectory that starts at $r_0$ at $t=0$, let us denote 
the position function with $r(r_0,t)$. We determine $r(r_0,t)$ by 
solving Eq.~(\ref{15}). The consistency between the trajectory 
and the wave function requires that
\begin{equation}
s(r_0) = \int_0 ^{r(r_0,t)}|\psi (r,t)|^2 \mathrm{d}r,
\label{19}
\end{equation}
at any time. This is based on the noncrossing nature of the 
trajectories.  In fact, Eq.~(\ref{19}) can be used as an 
alternative method of determining $r(r_0,t)$~\cite{8}. We have 
numerically solved Eq.~(\ref{15}) starting with a small but 
finite value of $t$, in order to avoid the highly oscillatory 
behavior of $\psi(r,t)$ as $t\rightarrow 0$. We have verified 
our solutions by varying the initial value of $t$ and also by 
using Eq.~(\ref{19}).

So far we have assumed that the trajectories start at $t=0$. 
We will assume the same in the rest of this Letter unless we
state otherwise. Let us note, however, that the starting time 
can be chosen at will. For a trajectory that starts at position 
$r$ and time $t>0$ with velocity $S'(r,t)$, one can unambiguously
trace its history back to $t=0$. We will do this later for the
trajectories shown in Fig.~\ref{fig3}.

The quantum decay process generally goes through three stages, 
I, II and III, which are characterized by different $t$ 
dependence of the nonescape probability $P(t)$ \cite{11,15},
\begin{equation}
P(t) = \int^a _0 |\psi(r,t)|^2 \mathrm{d}r.
\label{20}
\end{equation}
In the initial stage I we have, approximately, $1-P(t)\propto t^2$.
We have the exponential law $P(t)=\mathrm{e}^{-\Gamma t}$ in stage II
and the power law $P(t)\propto 1/t^3$ in final stage
III~\cite{16,17}\footnote{There is a claim that $P(t)\propto 1/t$ in
stage III \cite{16}. On the other hand it was argued that it is
$1/t^3$ \cite{17}. For our model we can calculate $P(t)$ directly by
using the explicit wave function of Eq.~(\ref{4}). We find that
$P(t)\propto 1/t^3$ in stage III, in agreement with \cite{17}.}.

Between the three stages are transition periods in which $P(t)$
exhibits an irregular, oscillatory behaviour~\cite{11}.  Stage II
usually spans most of the life time of the decaying system. In our
model with the chosen parameters the transition between stages I and
II is around $t=0.2$ and that between II and III is around
$t=12$. Toward the end of stage II, $P(t)$ becomes as small as
$10^{-8}$. By then the system has almost completely decayed. We
present results in five figures.

\begin{figure}[h]
\begin{center}
\epsfxsize4.in
\epsfbox{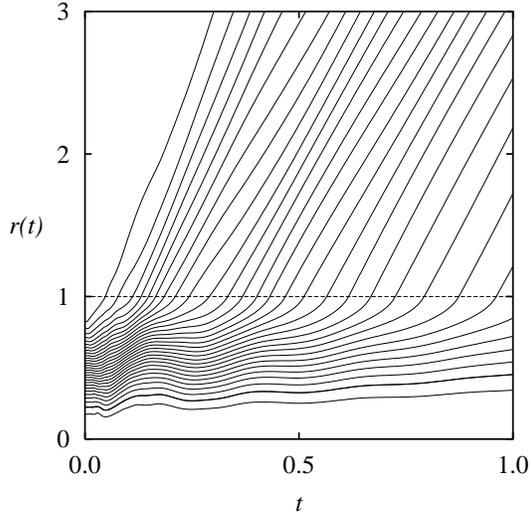}
\end{center}
\caption{Bohm trajectories (with $\lambda=6$) with equally
spaced $s$ of Eq.~(\ref{18}), with interval of $\Delta s
= 1/30$. These trajectories have equal statistical weight.
The trajectory density at $(r,t)$ is proportional to
the probability density $|\psi (r,t)|^2$.
The units are such that $\hbar=1$, $2m=1$ and $a=1$.}
\label{fig1}
\end{figure}

In Fig.~\ref{fig1} we show trajectories with equally spaced 
$s$, with interval of $\Delta s = 1/N$, $N=30$.  These 
trajectories have equal statistical weights. In other words, 
each of the trajectories occurs with the same probability. 
The trajectory density at $(r,t)$ is proportional to the 
probability density $|\psi (r,t)|^2$. The trajectories do not 
cross each other.  A trajectory that starts near the barrier 
($a=1$) escapes earlier.  Outside the barrier the trajectories 
become nearly straight. The slopes of the four or five 
trajectories that leave the barrier the earliest are somewhat
steeper than the others.
If we assume that $E(r,t)$ of Eq.~(\ref{17}) is conserved 
and that $U(r,t)$ is negligible for $r>a$, we obtain 
$E(r,t)=Q(r,0)=(\pi/a)^2$ for $r>a$. This leads to $p=\pi/a$ 
and $v=2\pi$ ($2m=1$, $a=1$) outside the barrier. This is
approximately the case as can be seen from the slopes of the 
trajectories of Fig. 1.

\begin{figure}[h]
\begin{center}
\epsfxsize3.4in
\epsfbox{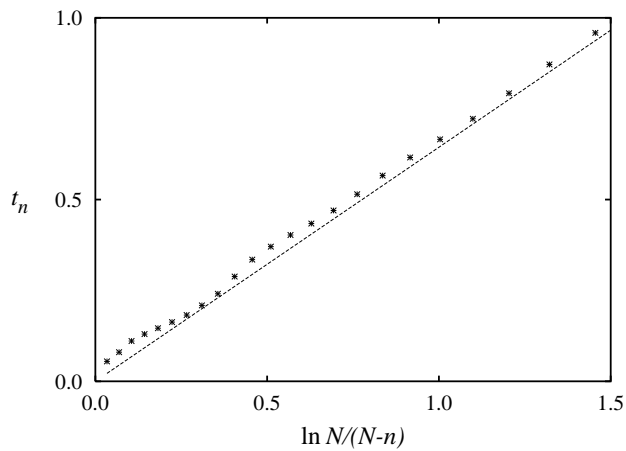}
\end{center}
\caption{
The escape time, at which the particle crosses the barrier at 
$r=a=1$ outward. The dots correspond to the trajectories of 
Fig.~\ref{fig1}. They are labeled with $n=1,2,\ldots \,$, 
starting with the one that escapes first. The dotted line is 
based on the exponential decay law, with $1/\Gamma = 0.644$. 
The parameters of the model and units are the same as in 
Fig.~\ref{fig1}. }
\label{fig2}
\end{figure}

Let us  label the trajectories with  $n=1,2, \ldots,29$, starting with
the  one that  escapes first.  In terms of  the starting  point $r_0$,
trajectory  1 is  the closest   to   the  barrier.  For  each of   the
trajectories we define the escape time $t_n$ as the  time at which the
particle crosses $a=1$ outward.  (Such  escape times or ``exit times''
have been formally discussed by Daumer et al.~\cite{daumer95} in
the context of the scattering problem in three dimensions.)  At $t=t_n$
out of the $N$ trajectories, $(N-n)$ trajectories are still within the
boundary. This means that the nonescape probability $P(t)$ is given by
\begin{equation}
P(t) =\frac {N-n}{N}, \hspace{0.1in}t_n <t<t_{n+1}. 
\label{21}
\end{equation}
Ideally $N$ should be taken as an infinitely large number. If the
exponential decay law holds exactly, that is, if $P(t) =
\mathrm{e}^{-\Gamma t}$, we obtain
\begin{equation}
t_n = \frac{1}{\Gamma}\ln \left( \frac{N}{N-n}\right),
\label{22}
\end{equation}
where $\Gamma$ is related to the half-life $\tau _{1/2}$ 
through $\tau_{1/2}=\ln 2/\Gamma$.  Figure~\ref{fig2} shows $t_n$
versus $\ln[N/(N-n)]$.  The dots correspond to the trajectories shown
in Fig.~\ref{fig1}. 

By fitting the numerically calculated nonescape probability of the
same model as ours with $P(t) = \mathrm{e}^{-\Gamma t}$, Winter obtained
$1/\Gamma = 0.644$ which leads to $\tau _{1/2}=0.446$~\cite{11}. In
Fig.~\ref{fig2} the dashed line shows the $n$ dependence of $t_n$
given by Eq.~(\ref{22}) with Winter's $1/\Gamma$. Except for the
first several ones, the dots follow the exponential curve very
well. The 15-th trajectory starts at $r_0=0.5$ and $s=0.5$.  It
crosses the barrier at $t_{15}= 0.468$, which is the half-life
$\tau_{1/2}$. This is slightly larger than the value 0.446 which is
based on Winter's estimate. (In Section 3 of \cite{13} below 
Eq.~(17) the half-life for $G=6$ was misquoted as $\tau_{1/2}=1.08$. 
The correct value is 0.446.)  Note that the decay process begins at 
a rate slower than predicted by the exponential law. This explains 
why the $\tau_{1/2}$ estimated by the trajectory of $s=1/2$ is 
greater than the one based on the exponential law.  

\begin{figure}[h]
\begin{center}
\epsfxsize3.4in
\epsfbox{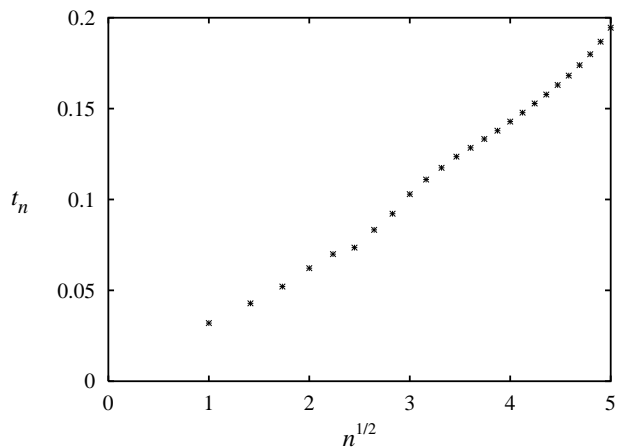}
\end{center}
\caption{The escape times at very small times.  In this case $N=100$
and $t_n$ is plotted against $\sqrt{n}$.
The parameters of the model and units are the same as in 
Fig.~\ref{fig1}. }
\label{fig2a}
\end{figure}

\begin{figure}[h]
\begin{center}
\epsfxsize4.in
\epsfbox{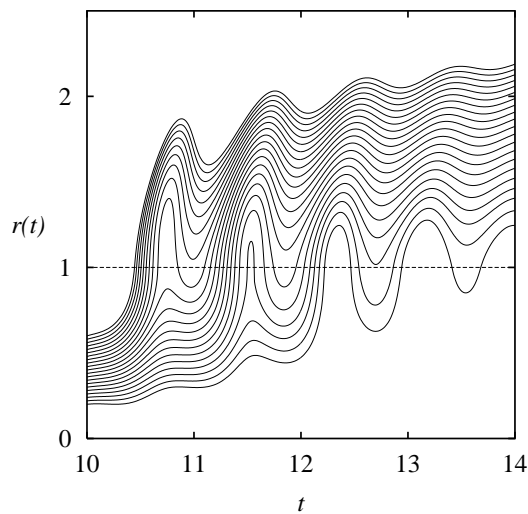}
\end{center}
\caption{Trajectories such that $r(t=10)$ ranges from 0.2
to 0.6 with the interval $\Delta r=0.02$. Because of this, 
unlike in Fig.~\ref{fig1}, the trajectory density is not 
proportional to the probability density. The parameters of
the model and units are the same as in Fig.~\ref{fig1}.}
\label{fig3}
\end{figure}

For $t<0.2$ the escape time does not follow the exponential law very
well.  It is in fact better fitted with $t_n\propto\sqrt{n}$.  In
Fig.~\ref{fig2a} we plot the escape time against $\sqrt{n}$.  In order
to see the details we have increased the number of trajectories, which
are again equally spaced with respect to $s$ but with a smaller
interval $\Delta s=1/N$ with $N=100$.  They are labelled with
$n=1,2,\ldots,99$ but the escape time is shown only for the first 25
trajectories.  The first three dots are almost exactly on a straight
line which passes through the origin.  This means that $1-P(t)\propto
t^2$ and hence $\mathrm{d}P(t)/\mathrm{d}t = 0$ at $t=0$.  In the
beginning of the decay process, $P(t)$ decreases more slowly than the
exponential law predicts.  This is analogous to the ``standby
mechanism'' which Elberfeld and Kleber~\cite{18} discussed in their
analysis of time-dependent tunneling of a semi-infinite wave train
through a thin barrier.  This deviation from the exponential law for
very small $t$ is a general feature of the quantum decay process which
is related to the possibility of the quantum Zeno effect as discussed
in~\cite{15}. For experimental evidence, see~\cite{19}. 

Let us take the model as a simulation of an $\alpha$-decaying 
nucleus and examine how the nuclear charge $Z(t)$ (in units of $e>0$) 
varies as a function of $t$ when the $\alpha$ particle of charge 2 
is emitted. In the traditional theory the charge number $Z(t)$ of 
the nucleus is given by
\begin{equation}
Z(t) = Z(0)- 2[1-P(t)].
\label{23}
\end{equation} 
The nonescape probability is well approximated by
$P(t) =e^{-\Gamma t}$ for most of the time. The $Z(t)$ changes 
from $Z(0)$ to $Z(0)-2$ gradually. This is because the wave 
function $\psi(r,t)$ of the $\alpha$ particle leaks out 
gradually. 

In the Bohmian description, if one follows the $n$-th 
trajectory, the nuclear charge changes from $Z(0)$ to 
$Z(0) -2$ suddenly at time $t_n$ when the $\alpha$ particle 
leaves the nucleus and hence
\begin{equation}
Z_n (t) = Z(0)-2\theta (t-t_n).
\label{24}
\end{equation} 
Here suffix $n$ refers to the $n$-th trajectory. In order
to obtain the nuclear charge that can be compared with that
of the traditional theory, we have to consider the ensemble of 
all trajectories each with a weight $|\psi(r_0,0)|^2$. This weight 
in the present case is $1/N$ for each trajectory. At time $t$ 
such that $t_n <t<t_{n+1}$, the particles of trajectories of 
1, 2, $\ldots , n$ have escaped. We thus obtain
\begin{equation}
Z(t) = Z(0) -\frac {2n}{N} = Z(0) -2[1-P(t_n)],
\label{25}
\end{equation} 
where we have used Eq. (\ref{21}). In the limit of 
$N \rightarrow \infty$, this $Z(t)$ converges to the $Z(t)$ 
of Eq. (\ref{23}) of the traditional theory. This illustrates
how a quantity that appears in the Bohmian description can
be related to its counterpart of the traditional theory.

Consider a gedanken experiment in which one tries to 
detect an $\alpha$ particle that is emitted from a source 
consisting of a single $\alpha$ emitting nucleus. The event 
in which one detects an $\alpha$ particle corresponds to one
of the Bohm trajectories. If one repeats this experiment
many times one experiences events, each of which is described
by one of the Bohm trajectories. Here it is understood
that the source is prepared every time in an identical manner.
In contrast to this, the Schr\"{o}dinger wave function does 
not describe any of the individual events, rather it only 
describes an ensemble of a large number of such events.
Instead of repeating the experiment on one system, we can 
think of experiments on many independent systems that are all 
identically prepared. In this sense, the word ``event(s)" 
can be replaced with ``system(s)".

Figure~\ref{fig3} shows 21 trajectories such that $r(t=10)$ ranges
from 0.2 to 0.6 with the interval $\Delta r=0.02$.  We have obtained
these trajectories by integrating Eq.~(\ref{15}) starting at
$t=10$. For the starting points, we have chosen to keep $\Delta r$
constant rather than $\Delta s$.  This choice is only a matter of
convenience or simplicity of the calculation involved. Because of this
choice, unlike in Fig.~\ref{fig1}, the trajectory density in this
figure is not proportional to the probability density. For example, in
the figure the statistical weight is larger for a trajectory with a
larger value of $r(t=10)$.  The range in terms of probability $s(r_0)$
is from $6.44\times 10^{-9}$ to $1.047\times 10^{-7}$.  As we stated
before we can easily trace the history of the trajectories back to
$t=0$. The values of $r_0$ at $t=0$ of the trajectories range from
$9.93\times 10^{-4}$ to $2.52\times 10^{-3}$. These trajectories are
starting almost from the origin. This is why they remain inside for a
very long time.

The trajectories go back and forth across the barrier. As Winter
pointed out many years ago, the current density $j(r,t)$ at the
barrier fluctuates in this time interval \cite{11,13}\footnote{There
are other space-time regions in which similar fluctuations (with
larger amplitudes) of the current density occur. See Fig. 5 of
\cite{13}.}.  At times it becomes negative, i.e.,
inward. Figure~\ref{fig3} visualizes this feature. In such a situation
we redefine the escape time as the time when the particle finally
leaves the barrier. The exponential law does not hold in this time
region any longer. Note also that, after leaking out through the
potential barrier, the trajectories tend to remain close to the
barrier. This situation is very different from that of
Fig.~\ref{fig1}. The time interval shown in Fig.~\ref{fig3} is the
transition period between stages II and III. In the latter, the
nonescape probability decreases like $1/t^3$.

\begin{figure}[h]
\begin{center}
\epsfxsize3.4in
\epsfbox{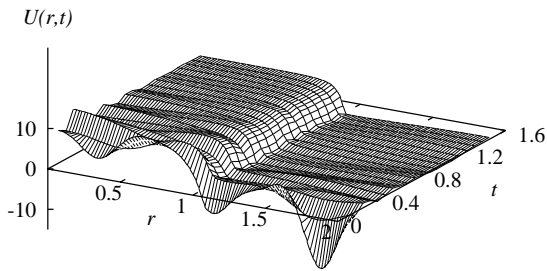}
\end{center}
\caption{Potential $U(r,t)$ which is equal to the quantum
potential $Q(r,t)$ with its $\delta$-function part removed.
The $U$ is continuous across the barrier $r=a\,(=1)$ except at
$t=0$. The $U(r,t)$ for $r<0.001$ and $t<0.05$ are not shown.
The parameters of the model and units are the same as in 
Fig.~\ref{fig1}.}
\label{fig4}
\end{figure}

Figure~\ref{fig4} shows the potential $U(r,t)$ which is equal to 
the quantum potential $Q(r,t)$ with its $\delta$-function part 
removed. There is no potential barrier in $U(r,t)$ and hence 
there is no tunneling phenomenon. As $r$ increases across the 
barrier at $r=a=1$, $U$ sharply drops but $U$ is continuous 
across the barrier (except at $t=0$). The behavior of $U$ is 
complicated for very small $t$ and also for very small $r$. The 
figure does not show the part of $t<0.05$ and $r<0.001$. Close 
scrutiny reveals that the trajectories rapidly fluctuate when 
$t$ and hence $r(t)$ are very small.  Although we do not show 
it, the behavior of $U(r,t)$ is also complicated in the 
space-time region that corresponds to Fig.~\ref{fig3}. 
For $r \gg a$, $U$ becomes negligible.

The results shown above are all for the case of $\lambda=6$. 
We have also examined the case of larger values of $\lambda$.  
For example, when $\lambda =100$, the deviations from the 
exponential law are very small. Let us add that, if we are 
to simulate $\alpha$ decay processes, we have to assume 
much larger values of $\lambda$, for example, of the order 
of 10$^8$ for $^{212}$Po. See Section 5 of \cite{13}. 

\section{Summary}
For the model defined by Eqs.~(\ref{1}) and (\ref{3}) we examined the
decay process from Bohm's point of view. We obtained Bohm trajectories
with which we can interpret various features of the decay process. We
see deviations from the exponential law at very small time and also at
very large time. The decay process is slower in the beginning than the
exponential law predicts.  In the time interval of $t=$ 10 to 14, we
showed that the trajectories go back and forth across the
barrier. This corresponds to the current density fluctuations that
Winter found. Beyond that time region, the exponential law is replaced
with a power law. One can verify this by showing that for large
times and very small $r_0$, $t_n\propto (N-n)^{-1/3}$, confirming that
the trajectories in this region of very large time are consistent with
$P(t) \propto 1/t^3$.

The results that we obtained in the Bohmian picture are
complementary but not contradictory to the traditional 
quantum mechanics. Let us, however, emphasize the following 
point. In contrast to the Schr\"{o}dinger wave function 
which describes an ensemble of a large number of events,
each of the Bohm trajectories represents an individual event. 
Such information on individual events is masked when the 
uncertainty regarding the starting points of the trajectories
is incorporated. In a situation in which one has to deal with 
an individual event, however, the Bohmian approach may lead 
to new insights. 

The model that we have considered is one of the simplest 
models for the decay process. The method that we have used 
can be applied to other potential models.

\ack
This work was supported by the Natural Sciences
and Engineering Research Council of Canada.

\end{document}